\documentclass{aa}
\usepackage[dvips]{graphicx}
\usepackage[mathscr]{eucal}
\usepackage{amssymb}
\usepackage{txfonts}

\newcommand{\be}{\begin{equation}}
\newcommand{\ee}{\end{equation}}
\newcommand{\bdm}{\begin{displaymath}}
\newcommand{\edm}{\end{displaymath}}

\begin{document}
   \title{High-energy emission of fast rotating white dwarfs}
\author{N.R.\,Ikhsanov\inst1\fnmsep\inst3 and
P.L.\,Biermann\inst2\fnmsep\inst4}
  \offprints{N.R.~Ikhsanov, Institute of Astronomy,
  University of Cambridge, Madingley Road, CB3 0HA, UK \\
  \email{ikhsanov@ast.cam.ac.uk}}

   \institute{Korea Astronomy and Space Science Institute, 61-1 Whaam-dong, Yusong-gu,
              Taejon 305-348, Republic of Korea
              \and
              Max-Planck-Institut f\"ur Radioastronomie, Auf dem
              H\"ugel 69, D-53121 Bonn, Germany
              \and
              Central Astronomical Observatory of the Russian
              Academy of Sciences, Pulkovo 65/1, 196140
              St.\,Petersburg, Russia
              \and
              Department of Physics and Astronomy, University of Bonn,
              Germany
              }

   \date{Received ; accepted }

\authorrunning{N.R.\,Ikhsanov and P.L.\,Biermann}
%\titlerunning{}

\abstract{The process of energy release in the magnetosphere of a
fast rotating, magnetized white dwarf can be explained in terms of
the canonical spin-powered pulsar model. Applying this model to the
white dwarf companion of the low mass close binary AE~Aquarii leads
us to the following conclusions. First, the system acts as an
accelerator of charged particles whose energy is limited to
$\mathcal{E}_{\rm p} \la 3$\,TeV and which are ejected from the
magnetosphere of the primary with the rate $L_{\rm
kin}\,\la\,10^{32}\,{\rm erg\,s^{-1}}$. Due to the curvature
radiation of the accelerated primary electrons the system should
appear as a source of soft $\gamma$-rays ($\sim 100$\,keV) with the
luminosity $< 3 \times 10^{27}\,{\rm erg\,s^{-1}}$. The TeV emission
of the system is dominated by the inverse Compton scattering of
optical photons on the ultrarelativistic electrons. The optical
photons are mainly contributed by the normal companion and the
stream of material flowing through the magnetosphere of the white
dwarf. The luminosity of the TeV source depends on the state of the
system (flaring/quiet) and is limited to $< 5 \times 10^{29}\,{\rm
erg\,s^{-1}}$. These results allow us to understand a lack of
success in searching for the high-energy emission of AE~Aqr with the
Compton Gamma-ray Observatory and the Whipple Observatory.

 }

   \maketitle
%%%%%%%%%%%%%%%%%%%%%%%%%%%%%%%%%%%%%%%%%%%%%%%%%%%%%%%%%

   \section{Introduction}

As shown by Usov (\cite{Usov-1988}), the process of energy release
in the magnetosphere of a fast rotating, magnetized white dwarf
can be explained in terms of a canonical spin-powered pulsar model
(e.g. Arons \& Scharlemann \cite{Arons-Scharlemann-1979}) provided
its surface temperature is $T(R_{\rm wd}) \la 10^6$\,K. The
appearance of such a white dwarf is similar to that of a
spin-powered neutron star (i.e. a canonical spin-powered pulsar)
in several important aspects. In particular, its energy budget is
dominated by the spin-down power
 \be
 L_{\rm sd} \simeq 3 \times 10^{35}\ \mu_{34}^2\
 \left(\frac{P_{\rm s}}{10\,{\rm s}}\right)^{-4}\ {\rm erg\,s^{-1}},
 \ee
which is spent mainly for the ejection of a relativistic wind.
Here $P_{\rm s}$ is the spin period of the white dwarf and
$\mu_{34}$ is its dipole magnetic moment expressed in units of
$10^{34}\,{\rm G\,cm^3}$. The pressure of the wind significantly
exceeds the ram pressure of the material surrounding the star at
the accretion (Bondi) radius and the particles accelerated in its
magnetosphere reach the TeV energies (Usov \cite{Usov-1988}).
According to Lipunov (\cite{Lipunov-1992}), the state of the white
dwarf can be classified as an {\it ejector}. Following this we
will refer such objects as Ejecting White Dwarfs (EWDs).

The relativistic particles accelerated in the magnetospheres of
EWDs manifest themselves in the high-energy part of the spectrum
(mainly due to the curvature radiation and the inverse Compton
scattering of thermal photons), while the dissipation of the
back-flowing current, which closes the current circuit in the
magnetosphere, leads to a local heating of the white dwarf surface
in the magnetic pole regions. Therefore, EWDs are expected to
appear as non-thermal $\gamma$-ray pulsars and pulsing X-ray/UV
sources with a predominantly thermal spectrum (see Usov
\cite{Usov-1993}).

A detailed analysis of the origin and appearance of EWDs has been
performed in the frame of the interpretation of non-identified
galactic $\gamma$-ray sources by Usov (\cite{Usov-1988}), and of
the anomalous X-ray pulsar 1E~2259+586 by Paczy\'nski
(\cite{Paczynski-1990}) and Usov (\cite{Usov-1993},
\cite{Usov-1994}). As they have shown, the basic appearance of
these objects fits well into the EWD-model and the formation of
fast rotating, magnetized white dwarfs should be a relatively
common phenomenon in our Galaxy. Nevertheless, further development
of these approaches was not effective mainly because of a lack of
evidence for the white dwarf nature of these sources (see e.g.
Mereghetti et~al. \cite{Mereghetti-etal-2002} and references
therein).

A target, to which the application of the EWD-model is currently
under consideration, is the degenerate component of the low mass
close binary system AE~Aqr. This star is unambiguously identified
with a fast rotating ($P_{\rm s}\simeq 33$\,s) white dwarf
(Eracleous et~al. \cite{Eracleous-etal-1994}) whose spin-down
power exceeds its luminosity by at least a factor of 120 (de~Jager
et~al. \cite{de-Jager-etal-1994}; Welsh \cite{Welsh-1999}). The
appearance of the system significantly differs from that of
magnetic cataclysmic variables whose radiation is powered by the
accretion process (Warner \cite{Warner-1995}). Instead, AE~Aqr
ejects material in the form of a non-relativistic stream (Welsh
et~al. \cite{Welsh-etal-1998}; Ikhsanov et~al.
\cite{Ikhsanov-etal-2004}), and relativistic particles responsible
for the radio and, possibly, high-energy emission of the system
(de~Jager \cite{de-Jager-1994}).

In this paper we analyze the appearance of AE~Aqr in the high
energy parts of the spectrum. The basic information on the system
and, in particular, its appearance in TeV $\gamma$-rays are
briefly outlined in Sect.\,2. The process of particle acceleration
in the magnetosphere of the white dwarf within the EWD-model is
discussed in Sect.\,3. In Sections\,4 and 5 we address the
appearance of the leptonic and hadronic components of the
accelerated particles, respectively. A brief summary and
discussion of acceleration mechanisms alternative to the EWD-model
are given in Sect.\,6.

   \section{AE~Aquarii}\label{aeaqr}

AE~Aqr is a non-eclipsing binary system with an orbital period of
9.88\,hours and eccentricity of 0.02, which is situated at the
distance of 100\,pc. The system inclination angle and the mass
ratio are limited to $48^{\degr} < i < 62^{\degr}$, and $0.58 \la
(q=M_2/M_1) \la 0.89$ (Welsh et~al. \cite{Welsh-etal-1995}).

The system emits detectable radiation in almost all parts of the
spectrum. It is a powerful non-thermal radio source (Bastian
et~al. \cite{bdch88}) and, possibly, a $\gamma$-ray emitter (see
below). Its optical, UV, and X-ray radiation is predominantly
thermal and comes from at least three different sites. The visual
light is dominated by the normal companion (secondary), which is a
K3--K5 red dwarf on or close to the main sequence (Welsh et~al.
\cite{Welsh-etal-1995}). The primary is a fast rotating,
magnetized white dwarf. The remaining light comes from a highly
variable extended source, which manifests itself in the blue/UV
continuum, the optical/UV broad single-peaked emission lines, and
the non-pulsing X-ray component. This source is associated with
the stream of material, which flows into the Roche lobe of the
white dwarf from the normal companion, interacts with the
magnetosphere of the primary and is ejected out from the system
without forming a disk (Wynn et~al. \cite{Wynn-etal-1997}; Welsh
et~al. \cite{Welsh-etal-1998}; Ikhsanov et~al.
\cite{Ikhsanov-etal-2004}). This source is also suspected of being
responsible for the peculiar rapid flaring of the star (Eracleous
\& Horne \cite{eh96}).

   \subsection{Parameters of the white dwarf}\label{wd-parameters}

A justification of the white dwarf nature of the primary has been
presented by Eracleous et~al. (\cite{Eracleous-etal-1994}) on the
basis of HST observations. As they have shown, the optical/UV
pulsing emission comes from two hot spots with the projected area
$A_{\rm p} \sim (0.8-4.3) \times 10^{16}\,{\rm cm^2}$ and the
temperature $20000$\,K$\,\la T_{\rm pol} \la 47000$\,K.
Associating these spots with the magnetic pole regions, they have
limited the angle between the rotational and magnetic axes of the
primary to $75^{\degr} \la \beta \la 77^{\degr}$ and evaluated the
temperature of the rest of its surface as 10000--16000\,K.

The white dwarf is spinning-down with a mean rate $\dot{P}_0 =
5.64 \times 10^{-14}\,{\rm s\,s^{-1}}$, which implies its
spin-down power (de~Jager et~al. \cite{de-Jager-1994}; Welsh
\cite{Welsh-1999})
   \be
L_{\rm sd} = 6 \times 10^{33}\ I_{50}\ \left(\frac{P_{\rm
s}}{33\,{\rm s}}\right)^{-3}\
\left(\frac{\dot{P}}{\dot{P}_0}\right)\ {\rm erg\,s^{-1}}.
  \ee
Here $I_{50}$ is the moment of inertia of the white dwarf
expressed in units of $10^{50}\,{\rm g\,cm^2}$.

$L_{\rm sd}$ exceeds the luminosity of the system observed in the
UV and X-rays by a factor of 120 and its persistent bolometric
luminosity by a factor of 5. As mentioned above, such a situation
is typical for EWDs. The dipole magnetic moment of the white dwarf
within this approach can be evaluated as (Ikhsanov
\cite{Ikhsanov-1998})
   \be\label{magmom}
\mu_0 \simeq 1.4 \times 10^{34}\ \left(\frac{P_{\rm s}}{33\,{\rm
s}}\right)^{2} \left(\frac{L_{\rm sd}}{6 \times 10^{33}\,{\rm
erg\,s^{-1}}}\right)^{1/2} {\rm G\,cm^{3}}.
   \ee
This implies that the strength of the surface magnetic field of
the white dwarf in the magnetic pole regions is (Alfv\'en \&
F\"althammar \cite{af63})
 \be\label{B0}
 B_0 = \frac{2 \mu}{R_{\rm wd}^3} \simeq 100\ R_{8.8}^{-3}\
 \left[\frac{\mu}{1.4 \times 10^{34}\,{\rm G~cm^3}}\right]\ {\rm MG},
 \ee
and, correspondingly, the surface field strength at its magnetic
equator is $B_0/2 = 50$\,MG. Here $R_{8.8}$ is the radius of the
white dwarf expressed in units of $10^{8.8}$\,cm.

As recently shown by Ikhsanov et~al. (\cite{Ikhsanov-etal-2004}),
the above mentioned estimates contradict none of the currently
observed properties of the system, and the Doppler H$\alpha$
tomogram simulated within this approach is in a very good
agreement with the tomogram derived from spectroscopic
observations of AE~Aqr. On this basis estimates~(\ref{magmom}) and
(\ref{B0}) will be used in our analysis.

  \subsection{High-energy emission}\label{tev-obs}

A search for TeV emission of AE~Aqr was performed by three
independent groups. The Potchefstroom group has reported on about
310\,hours of observations in 1988--93 (Meintjes et~al.
\cite{Meintjes-etal-1992}; \cite{Meintjes-etal-1994}). A few
events of pulsing TeV emission with an averaged flux $(0.2-2)
\times 10^{-10} {\rm cm^{-2}\,s^{-1}}$ (at a threshold energy
2.4\,TeV) were detected during this period. Additionally, they
reported a detection of two short (1 and 3\,minutes duration)
unpulsed outbursts on 25 June 1993 (the orbital phases 0.04 and
0.05). The flux of the TeV emission during these events was about
$8 \times 10^{-9} {\rm cm^{-2}\,s^{-1}}$, which, under the
assumption of isotropic emission, corresponds to the source
luminosity of $\sim 10^{34}\,d_{100}^2\,{\rm erg\,s^{-1}}$. Here
$d_{100}$ is the distance to AE~Aqr normalized to 100\,pc.

A detection of TeV emission from AE~Aqr has also been reported by
the Durham group (Bowden et~al. \cite{Bowden-etal-1992}; Chadwick
et~al. \cite{Chadwick-etal-1995}). They recorded about 170\,hours
of data in 1990--93. A persistent component of pulsing (at
60.46\,mHz) TeV emission with a time-averaged flux of about
$10^{-10}\,{\rm cm^{-2}\,s^{-1}}$ (at energy $>\,350$\,GeV) was
detected. Additionally, they reported the detection of a 1\,minute
(13 October 1990, orbital phase 0.40) and a 70\,minutes (11
October 1993, orbital phases 0.62--0.74) outburst. The luminosity
of the $\gamma$-ray source during these outbursts was $\sim 6
\times 10^{33}\,d_{100}^2\,{\rm erg\,s^{-1}}$ and $3 \times
10^{32}\,d_{100}^2\,{\rm erg\,s^{-1}}$, respectively. In both
cases the signal was modulated with a half of the spin period of
the white dwarf (60.46\,mHz).

On the other hand, the analysis of data recorded by the Whipple
group (68\,hours of observations in 1991--95) have shown no
evidence for any steady, pulsed or episodic TeV emission of AE~Aqr
(Lang et~al. \cite{Lang-etal-1998}). The upper limits to the flux
of a steady emission, $\la 4.0 \times 10^{-12}\,{\rm
cm^{-2}\,s^{-1}}$, and a pulsing emission $\la 1.5 \times
10^{-12}\,{\rm cm^{-2}\,s^{-1}}$, at an energy threshold of
900\,GeV have been derived. This suggests that the persistent
luminosity AE~Aqr in TeV energy range is unlikely to exceed $5
\times 10^{30}\,{\rm erg\,s^{-1}}$. At the same time, the database
of the Whipple group is too small for any conclusions on the
non-frequent TeV $\gamma$-ray events reported by the Potchefstroom
and Durham groups (for a discussion see Lang et~al.
\cite{Lang-etal-1998}).

Finally, a negative result of a search for 0.1--1\,GeV emission
from AE~Aqr with the Compton Gamma-Ray Observatory (CGRO) has been
reported by Schlegel et~al. (\cite{Schlegel-etal-1995}). The
derived upper limits suggest that the luminosity of the system in
the EGRET energy range is smaller than $10^{31}\,{\rm
erg\,s^{-1}}$.

A lack of success in searching for $\gamma$-ray emission from AE~Aqr
reported by the EGRET and Whipple groups raises the following
questions: Do we have any theoretical grounds to suggest that this
system can be an emitter of high-energy radiation and particles and
if so, how bright this source could be\,? In order to answer these
questions we performed an analysis of particle acceleration based on
the EWD-model of the system. We show that the expected intensity of
high-energy emission within this model is indeed below the
thre\-shold of detectors used by the CGRO and Whipple observatories.

   \section{Particle acceleration}\label{particle-acceleration}

We consider a process of particle acceleration in the
magnetosphere of a white dwarf, whose rotation period $P_{\rm
s}=33\,P_{33}$\,s (the angular velocity $\Omega=2\pi/P_{\rm s}
\simeq 0.19\,{\rm rad\,s^{-1}}$), the mass $M_{\rm
wd}=0.8\,M_{0.8}\,M_{\sun}$, the strength of the magnetic field at
the magnetic pole regions $B_0=10^8\,B_8$\,G and the surface
temperature $T < 50000$\,K. Under these conditions a scale height
of its atmosphere,
 \be
 h_{\rm a} \simeq \frac{k T R_{\rm wd}^2}{m_{\rm p} GM_{\rm wd}}
 \simeq 2.5 \times 10^4\ M_{0.8}^{-1}\ R_{8.8}^2\
 \left(\frac{T}{5\times 10^4\,{\rm K}}\right)\ {\rm cm},
 \ee
is significantly smaller than both the radius of the white dwarf
($R_{\rm wd} \simeq 10^{8.8}$\,cm) and the lower limit to the
radius of its polar caps,
 \be
\Delta R_{\rm p} \simeq \left(\frac{\Omega R_{\rm
wd}}{c}\right)^{1/2} R_{\rm wd} \simeq 4 \times 10^7\
R_{8.8}^{3/2}\ \left(\frac{P_{\rm s}}{33\,{\rm s}}\right)^{-1/2}\
{\rm cm}.
 \ee
This implies that the vacuum approximation to the medium
surrounding the white dwarf is applicable and therefore, the
process of particle acceleration can be treated in terms of
EWD-model.

According to Arons \& Scharlemann (\cite{Arons-Scharlemann-1979}),
the component of the electric field $E_{\parallel} \equiv(\vec{E}
\cdot \vec{B})/|\vec{B}|$ along the magnetic field $\vec{B}$,
which is generated in the polar cap regions of a fast rotating
magnetized star surrounded by a vacuum can be evaluated as
 \be
 E_{\parallel} \simeq E_{\rm AS} \times \left\{
\begin{array}{ll}
s/\Delta R_{\rm p} & \hspace{.2cm}
\mbox{at } 0 \leq s \leq \Delta R_{\rm p},\\
1 & \hspace{.2cm} \mbox{at } \Delta R_{\rm p} \leq s \leq R_{\rm wd}, \\
\sqrt{2R_{\rm wd}/r}  & \hspace{.2cm} \mbox{at } s > R_{\rm wd},
\end{array}
\right.
  \ee
where
 \be\label{eas}
 E_{\rm AS} = \frac{1}{8\sqrt{3}} \left(\frac{\Omega R_{\rm
 wd}}{c}\right)^{5/2} B_0,
 \ee
and $r=(R_{\rm wd}+s)$ is a distance from the surface of the white
dwarf. This means that the electric potential in the magnetosphere
of the white dwarf is
 \be\label{phi}
 \varphi_{\rm as}(r) = \int_{\rm R_{\rm wd}}^{\rm r} E_{\parallel}\ ds\ \simeq\
 2 \sqrt{2}\ E_{\rm AS}\ R_{\rm wd} \left[\left(\frac{r}{R_{\rm
wd}}\right)^{1/2} - 1\right].
 \ee

The total electric potential in the magnetosphere of a spin-powered
neutron star contains two additional terms which represent the
effects of general relativity (Muslimov \& Tsygan
\cite{Muslimov-Tsygan-1992}), and the effect of $e^{\pm}$ pairs
creation in the outer magnetosphere (Cheng et~al.
\cite{Cheng-etal-1986}). However, a contribution of these terms to
the electric potential in the magnetosphere of a white dwarf (whose
radius exceeds that of a neutron star by a factor of 500 and the
magnetic field is too weak for a pair creation process to be
effective) are small and can be neglected.

An application of the EWD-model to the white dwarf in AE~Aqr implies
a validity of the following two assumptions. The first is a lack of
accretion of material onto the surface of the white dwarf. If this
assumption were not valid the star would appear as an
accretion-powered X-ray pulsar. The system is indeed an emitter of
X-rays. But all of currently established properties of the emission
are inconsistent with those predicted by the accretion-powered
pulsar model. In particular, the X-ray spectrum of AE~Aqr is soft
and significantly differs from the typical spectra of
accretion-powered white dwarfs (Clayton \& Osborne
\cite{Clayton-Osborne-1995}; Choi et~al. \cite{Choi-etal-1999}).
Furthermore, the surface temperature of the white dwarf at the
magnetic pole regions evaluated by Eracleous et~al.
(\cite{Eracleous-etal-1994}) is $<50000$\,K, i.e. a factor of 2000
smaller than the typical temperature at the base of an accretion
column. Finally, observations of AE~Aqr with XMM-Newton recently
reported by Itoh et~al. (\cite{Itoh-etal-2005}) suggest that the
number density of plasma responsible for the detected X-rays is
$n_{\rm e}\sim 10^{11}\,{\rm cm^{-3}}$ (i.e. a few orders of
magnitude lower than corresponding conventional estimates in the
post-shock accretion column) and that the linear scale of the source
is $\ell_{\rm p} \ga 2 \times 10^{10}$\,cm (i.e. a factor of 40
larger than the radius of the white dwarf). This clearly indicates
that the source of the observed X-ray emission is located at a
significant distance from the white dwarf and therefore, cannot be
powered by an accretion of material onto its surface.

The second assumption is that the material streaming through the
magnetosphere of the white dwarf does not significantly effect the
electric potential generated at its surface. A hint for a validity
of this assumption comes from the model of Michel \& Dessler
(\cite{Michel-Dessler-1981}) who have considered a situation in
which a spin-powered pulsar is surrounded by a dead disk. According
to their results a presence of a disk inside the light cylinder of
the pulsar would not suppress the electric potential at the stellar
surface. Instead, an interaction between the field lines and the
disk material leads to an additional term in the expression of the
total electric potential at the stellar surface. This finding has
already been used by de~Jager (\cite{de-Jager-1994}) for
constructing a dead disk model for AE~Aqr. The value of the electric
potential in this case is comparable to that of $\varphi_{\rm
as}(r_0)$, where $r_0$ is the radius of the disk (or a distance of
closest approach of the stream to the white dwarf). In this light
expression~(\ref{phi}) represents an upper limit to the electric
potential which could be generated in the magnetosphere of the white
dwarf within the EWD-model.

The energy of particles accelerated in the $\varphi_{\rm as}(r)$
potential can be limited to $\mathcal{E}_{\rm p} \la
\mathcal{E}_0$, where
 \be\label{emax}
\mathcal{E}_0 \simeq 2 \times 10^{11}\ \mu_{34.2}\ R_{8.8}^{1/2}\
\left(\frac{P_{\rm s}}{33\,{\rm s}}\right)^{-5/2}\
\left[\left(\frac{l_0}{R_{\rm wd}}\right)^{1/2} - 1\right]\ {\rm
eV}.
 \ee
Here $l_0$ is the scale of the acceleration region. A precise
determination of this parameter is an open problem so far. In
general case the value of $l_0$ can be limited to (e.g. Beskin
et~al. \cite{Beskin-etal-1993}) $(R_{\rm wd}+\Delta R_{\rm p})
\leq l_0 \leq R_{\rm lc}$, where $R_{\rm lc} = c/\Omega \simeq 1.6
\times 10^{11}\ (P/33\,{\rm s})$\,cm is the radius of the light
cylinder of the white dwarf. Following this limitation one can
estimate the energy of the accelerated particles as
 \be
8\,{\rm GeV} \la \mathcal{E}_{\rm p} \la 3\,{\rm TeV}.
 \ee

The kinetic luminosity of the particle beam can be evaluated as
 \be\label{lkin}
 L_{\rm kin} \simeq e \varphi_{\rm as}(l_0) \dot{N},
 \ee
where
 \be\label{ndot}
 \dot{N} = \pi (\Delta  R_{\rm p})^2 n_{\rm GJ}(R_{\rm wd}) c
 \ee
is the flux of ultrarelativistic particles from the white dwarf
and
 \be\label{ngj}
n_{\rm GJ}(R_{\rm wd}) = \frac{(\vec{\Omega} \cdot \vec{B})}{2 \pi c
e} \simeq 5 \times 10^4\ \left(\frac{P_{\rm s}}{33\,{\rm
s}}\right)^{-1}\ \left(\frac{B_0}{10^8\,{\rm G}}\right)\ {\rm
cm^{-3}}
 \ee
is the Goldreich-Julian density above the primary surface.

Combining Eqs.~(\ref{eas}) and (\ref{phi}) with
Eqs.~(\ref{lkin})--(\ref{ngj}) and setting $l_0 = R_{\rm lc}$
yields
 \be
 L_{\rm kin}^{\rm max} = \frac{1}{4 \sqrt{6}}\ \frac{\Omega^4
 R_{\rm wd}^6 B_0^2}{c^3} \cos{\beta}\ \simeq\ 0.04 L_{\rm sd},
 \ee
where it is taken into account that
 \be
 L_{\rm sd} = \frac{2}{3} \frac{\Omega^4
 R_{\rm wd}^6 B_0^2}{c^3}.
 \ee
Thus, the maximum kinetic luminosity of the beam of
ultrarelativistic particles ejected by the white dwarf in AE~Aqr
is
 \be
 L_{\rm kin}^{\rm max} \simeq 2 \times
 10^{32}\ \left(\frac{L_{\rm sd}}{6\times 10^{33}\,{\rm erg\,s^{-1}}}\right)\
 {\rm erg\,s^{-1}}.
 \ee
It is comparable to the bolometric luminosity of the system during
its optical quiescent state (see Beskrovnaya et~al.
\cite{Beskrovnaya-etal-1996}).

If, however, the scale of the acceleration region is close to its
minimum value, i.e. $l_0 \sim (R_{\rm wd}+\Delta R_{\rm p})$, the
kinetic luminosity of the beam of GeV-electrons does not exceed
  \be
L_{\rm kin}^{\rm min} \simeq 5 \times
 10^{29}\ \left(\frac{L_{\rm sd}}{6\times 10^{33}\,{\rm erg\,s^{-1}}}\right)\
 {\rm erg\,s^{-1}}.
 \ee

  \section{Radiative losses of primary electrons}

The radiative losses of ultrarelativistic electrons accelerated in
the magnetosphere of a magnetized compact star are governed mainly
by two mechanisms: i)~the curvature radiation, and ii)~the inverse
Compton scattering of thermal photons (Michel \cite{Michel-1991}).

   \subsection{Curvature radiation}

The intensity and the mean photon energy of the curvature
radiation emitted by ultrarelativistic electrons in the
magnetosphere of the white dwarf are (Usov \cite{Usov-1988})
 \be\label{lc}
 l_{\rm c} = \frac{2}{3} \frac{e^2 c}{R_{\rm c}^2} \gamma^4,
 \ee
 \be\label{ecurv}
 \bar{E}_{\rm curv} = \frac{3}{2}\ \frac{\hbar c \gamma^3}{R_{\rm
 c}}.
 \ee
Here $R_{\rm c} \simeq (Rc/\Omega)^{1/2}$ is the curvature radius
of the magnetospheric field lines and $\gamma = \mathcal{E}_{\rm
p}/mc^2$ is the Lorentz factor of electrons which in the case of
AE~Aqr can be expressed using Eq.~(\ref{emax}) as
 \be\label{gammar}
 \gamma_{\rm e}(r) \simeq 4 \times 10^5\ \mu_{34.2}\ \left(\frac{P_{\rm
s}}{33\,{\rm s}}\right)^{-5/2}\
 \left[\left(\frac{r}{R_{\rm wd}}\right)^{1/2} - 1\right].
 \ee
Since the Lorentz factor of the electrons and the values of the
parameters $l_{\rm c}(r)$ and $\bar{E}_{\rm curv}(r)$ increase
with r, the largest fraction of the curvature radiation is
generated near the light cylinder of the white dwarf. Therefore,
the upper limit to the mean photon energy and the intensity of the
curvature radiation ($L_{\rm curv} \simeq l_{\rm c} \dot{N}$)
expected from AE~Aqr can be evaluated from Eqs.~(\ref{lc}) --
(\ref{gammar}) as
 \bdm
\gamma_{\rm e}(R_{\rm lc}) \simeq 6 \times 10^6
 \edm
 \bdm
\bar{E}_{\rm curv}(R_{\rm lc}) \simeq 100\ {\rm keV}
 \edm
 \bdm
 L_{\rm curv}(R_{\rm lc}) \la 3 \times 10^{27} \mu_{\rm 34.2}^5\
 \left(\frac{P_{\rm s}}{33\,{\rm s}}\right)^{-11}\ {\rm erg\,s^{-1}}.
 \edm
Hence, the curvature radiation of the primary electrons
contributes to the system emission mainly in the soft ($\sim
100$\,keV) $\gamma$-rays with a mean flux of $\sim 10^{-15}\,{\rm
erg\,cm^{-2}\,s^{-1}}$. This value is significantly smaller than
the threshold sensitivity of detectors used in the EGRET
experiment. This provide us with a reasonable explanation of
unsuccessful search for high-energy emission of the system with
CGRO.

   \subsection{Inverse Compton scattering}

The mechanism, which can be responsible for the very high-energy
emission of the system, is the inverse Compton scattering of the
thermal photons on the ultrarelativistic electrons. The energy of
the scattered photons is as high as $m_{\rm e}c^2 \gamma (R_{\rm
lc}) \sim 3$\,TeV, where $m_{\rm e}$ is the electron mass.

The intensity of the TeV radiation generated due to the inverse
Compton scattering is (Ochelkov \& Usov \cite{Ochelkov-Usov-1983},
the Klein-Nishina cross-section case)
 \be\label{lvhe}
 L_{\rm cs} \simeq l_{\rm cs} \dot{N},
 \ee
where
 \be\label{lcs}
 l_{\rm cs} = \frac{3}{8} c \sigma_{\rm T} u_0 \left(\frac{m_{\rm
e}c^2}{\epsilon_{\rm th}}\right)^2 \left[\ln{\frac{4 \epsilon_{\rm
th} \gamma}{m_{\rm e}c^2}} -
 \frac{11}{6}\right]
 \ee
is the rate of energy losses of an electron. Here $\epsilon_{\rm
th} \simeq 3kT$ is the mean energy of the thermal photons,
$\sigma_{\rm T}$ is the Thomson cross section, $k$ is the
Boltzmann constant, $T$ is the mean temperature of the thermal
emission and
 \be\label{u0}
 u_0 \simeq a T^4 \left(\frac{R_0}{r_{\rm s}}\right)^2
 \ee
is the radiation energy density. $R_0$ denotes the effective
radius of the source of the thermal emission, $r_{\rm s}$ is the
mean distance from this source to the region of scattering and $a
= 7.6 \times 10^{-15}\,{\rm erg\,cm^{-3}\,deg^{-4}}$ is the
radiation density constant.
% The Eq.~(\ref{lcs}) is written for the
% case of the quantum Compton scattering (i.e. the Klein-Nishina
% scattering cross section has been used) which is appropriate in
% the case considered.

The field of the thermal radiation in the magnetosphere of the white
dwarf in the case of AE~Aqr is contributed by three separate
sources: the normal companion, the white dwarf and the stream of
material flowing through its magnetosphere. The relative
contribution of these sources depends on the location of the
scattering region and the state of the system (quiescent or
flaring). As we require the energy of the scattered photons to be
$\sim$\,TeV, the closest distance of the region of their generation
to the surface of the white dwarf is $r \ga 2 \times 10^{10}$\,cm
(see Eq.~\ref{gammar}). The contribution of the white dwarf to the
radiation field at this distance is significantly smaller than those
of the normal component and of the stream. Furthermore, the ratio
$\eta = l_{\rm cs}/eE_{\parallel}c$ in the vicinity of the white
dwarf is
 \bdm
 \eta \simeq 10^{-4} \left(\frac{T}{10^4\,{\rm K}}\right)^2
 \left(\frac{R_{\rm wd}}{r}\right)^{3/2}.
 \edm
This indicates that the thermal emission from the surface of the
white dwarf is not essential for the process of particle
acceleration. On this basis we neglect the contribution of the white
dwarf in further consideration.

As recently shown by Ikhsanov et~al. (\cite{Ikhsanov-etal-2004}),
the closest approach of the stream to the surface of the white
dwarf within the EWD-model is limited to
  \be\label{ralf}
r_0 \ga 4 \times 10^{10}\ \eta_{0.37}\ \mu_{34.2}^{4/7}\
M_{0.8}^{-1/7}\ \dot{M}_{17}^{-2/7}\ {\rm cm},
 \ee
where $\dot{M}_{17}$ is the mass transfer rate expressed in units
of $10^{17}\,{\rm g\,s^{-1}}$ and $\eta_{0.37}=\eta/0.37$ is the
parameter accounting for the geometry of the accretion flow, which
in the case of a stream is normalized following Hameury et~al.
(\cite{hkl86}). This indicates that the contribution of the stream
to the radiation field during the quiescent state ($\dot{M} <
10^{17}\,{\rm g\,s^{-1}}$) is comparable to that of the normal
companion inside the magnetosphere and is almost negligible at the
distances close to the light cylinder ($R_{\rm lc} = 1.57 \times
10^{11}$\,cm). Therefore, using the parameters of the secondary
(i.e. $T_{\rm eff} \simeq 4500$\,K, $R_0 \simeq R_{\sun}$) one can
evaluate the intensity of the TeV emission produced by the inverse
Compton scattering during the quiescent state of the system as
(see Eqs.~\ref{lvhe}--\ref{u0})
 \be
 L_{\rm cs}^{\rm q} \la 3 \times 10^{27}\ {\rm erg\,s^{-1}}.
 \ee
Here we assumed that the scattering occurs at the light cylinder
of the primary.

The most favorable conditions for the generation of TeV emission
due to the inverse Compton scattering occur during the flaring
state. According to Beskrovnaya et~al.
(\cite{Beskrovnaya-etal-1996}), the system emission during the
strongest optical flares (which last a few minutes) is dominated
by a source with an effective temperature $\sim 20000$\,K and an
effective area of $\simeq 10^{20}\,{\rm cm^2}$. As this source is
situated inside the magnetosphere (at the distances of about
$r_0$) its contribution to the radiation field significantly
exceeds that of the normal companion (the optical luminosity of
the system during these events reaches $\simeq 10^{33}\,{\rm
erg\,s^{-1}}$). In this case the generation of TeV emission due to
the inverse Compton scattering reaches its highest efficiency in a
region situated at the distances of about $r_0$ from the surface
of the primary and its intensity (according to Eqs.~\ref{lvhe} --
\ref{u0}) can be limited to
 \be
 L_{\rm cs}^{\rm f} \la 4 \times 10^{29}\,{\rm erg\,s^{-1}}
 \ee
(the mass transfer rate during the strong flares according to the
EWD-model is $\sim 5 \times 10^{17}\,{\rm g\,s^{-1}}$).

Thus, the upper limit to the intensity of TeV radiation of AE~Aqr
produced due to the inverse Compton scattering of thermal photons
on the ultrarelativistic electrons is a factor of a few smaller
the threshold sensitivity of detectors used in observations of the
system with the Whipple $\gamma$-ray telescope. On the other hand
the evaluated luminosity of the source is high enough for its
emission to be detected in the MAGIC, VERITAS and HESS
experiments.

   \section{Appearance of a proton beam}

One can also envisage a situation in which a significant fraction
of $L_{\rm sd}$ is converted to ultrarelativistic protons. The
upper limit to the kinetic luminosity of the proton beam in this
case is expressed by Eq.~(\ref{lkin}). The radiative losses of the
protons in the magnetosphere of the white dwarf (due to the
curvature radiation and the inverse Compton scattering) are
significantly smaller than those of electrons and in the first
approximation can be neglected. The energy of the beam, however,
can be effectively converted into VHE $\gamma$-rays if the
trajectories of protons intersect a target of a relatively dense
($30-80\,{\rm cm^{-2}}$) background material. In this case a
creation of $\pi$-mesons ($\pi^0$, $\pi^{\pm}$) and their
subsequent decays into two VHE $\gamma$-photons with the energy of
about the energy of the primary relativistic protons would be
expected.

Two targets with the required column density can be indicated in
AE~Aqr. Namely, the atmosphere of the normal companion, and the
stream of material moving through the magnetosphere of the white
dwarf. The interaction of TeV protons with these targets will
produce a flux of TeV photons. However, can this emission be
detected by an observer at the Earth\,? A positive answer on this
question in the light of currently established geometry of the
system is not obvious. Indeed, the photons generated in this
process are strongly beamed along the incident particle's velocity
vector (see e.g. Vestrand \& Eichler
\cite{Vestrand-Eichler-1982}). Therefore, an observer can detect
only those photons that are produced when protons streaming toward
him strike intervening target material. This means that the flux
of the TeV photons generated in the atmosphere of the secondary
can be detected only at the orbital phases $(2 \pi - \Delta
\varphi) \la \varphi \la (2 \pi +\Delta \varphi)$, where
 \be\label{deltavarphi}
\Delta \varphi = \frac{1}{2 \pi} \arctan{\left(\frac{R_{\rm
L(2)}}{A}\right)} = \frac{1}{2 \pi}
\arctan{\left(0.378\,q^{-0.2084}\right)},
 \ee
and if the inclination angle of the system is
 \be\label{i0}
i \ga i_0 \simeq \frac{\pi}{2} -
\arctan{\left(0.378\,q^{-0.2084}\right)}.
 \ee
Here $R_{\rm L(2)}$ is the volume radius of the Roche lobe of the
secondary and $A$ is the orbital separation. For the parameters of
AE~Aqr conditions~(\ref{deltavarphi}) and (\ref{i0}) are $0.94 \la
\varphi \la 1.06$ and $i > i_0 \sim 67^{\degr}-69^{\degr}$,
respectively. Hence the atmosphere of the secondary is definitely
not a site of TeV emission detected at orbital phases $0.07 -
0.93$. Furthermore, it is unlikely a site of the TeV emission at
the rest of the orbital phases since the assumption $i
> 67^{\degr}$ contradicts optical observations of the system (see
Sect.\,\ref{aeaqr}).

The TeV emission generated due to the interaction between the
relativistic protons and the stream could be observed in a
significantly wider range of orbital phases. However, the
condition to the inclination angle of the system in this case is
  \be\label{istream0}
 i > i_0 \simeq \frac{\pi}{2} - \arctan{\left(\frac{r_{\rm
str}}{r_0}\right)},
 \ee
where $r_{\rm str}$ is the radius of the stream cross section,
which in the considered case is limited to (see e.g. Wynn et~al.
\cite{Wynn-etal-1997}; Ikhsanov et~al. \cite{Ikhsanov-etal-2004}):
 \be\label{rhos}
r_{\rm str} \la (Q_{\rm L1}/\pi)^{1/2} \simeq 1.6 \times 10^{9}\
\left(\frac{T_{\rm eff}}{4500\,{\rm K}}\right)^{1/2} \ {\rm cm},
 \ee
and $r_0$ is the distance of closest approach of the stream to the
white dwarf, which in the case of EWD-model is expressed by
Eq.~(\ref{ralf}). Here $Q_{\rm L1}$ is the effective cross section
of the mass transfer throat at the Lagrange point L1 and $T_{\rm
eff}$ is the effective temperature of the photosphere of the
secondary. Combining Eqs.~(\ref{ralf}), (\ref{istream0}) and
(\ref{rhos}) one finds that the TeV photons generated due to
interaction between the accelerated protons and the stream could
be detected by an observer if the inclination angle of the system
were $\ga 86^{\degr}$. This, however, is inconsistent with the
value of $i$ derived from observations (otherwise the system would
be an eclipsing binary).

Finally, we would like to point out that in case of any targets a
strong collimation of the beam of ultrarelativistic protons  is
required for the effective production of TeV emission. Indeed, the
reported luminosity of the TeV source during outbursts is an order
of magnitude larger than the upper limit to the kinetic luminosity
of the beam evaluated in Sect.\,\ref{particle-acceleration}.
Taking into account that the efficiency of the energy transfer
from protons to $\pi^0$-mesons does not exceed 20\% (Ozernoy
et~al. \cite{Ozernoy-etal-1973}), and that only a half of the
energy of the $\pi^0$-mesons is transferred to $\gamma$-rays
detected by an observer one can limit the opening angle of the
beam to
 \be
 \Delta \Upsilon \la 0.1 \varkappa \left(\frac{L_{\rm kin}}{9
 \times 10^{32}\,{\rm erg\,s^{-1}}}\right) \left(\frac{L_{\rm obs}}
 {10^{34}\,{\rm erg\,s^{-1}}}\right) {\rm rad},
 \ee
where
 \be
 \varkappa = \left\{
\begin{array}{ll}
1 & \hspace{.2cm} \mbox{for } \hspace{.2cm} Q_{\rm b} \la Q_{\rm t},\\
Q_{\rm t}/Q_{\rm b} & \hspace{.2cm} \mbox{for } \hspace{.2cm} Q_{\rm b} > Q_{\rm t}. \\
\end{array}
\right.
  \ee
Here $Q_{\rm b}$ and $Q_{\rm t}$ are the cross sections of the
beam and the target in the direction, perpendicular to the
velocity vector of the relativistic protons. This means that the
opening angle of the beam for any targets should be less than
$7^{\degr}$. This value corresponds to the opening angle of the
dipole magnetic field at the distance $r_1 \simeq 2.5\,R_{\rm wd}$
from the surface of the primary. Therefore, to meet this criterium
one has to assume that either a target is situated very close to
the surface of the white dwarf or the accelerated protons at $r >
r_1$ do not follow the magnetic field lines but propagate through
the magnetosphere along almost straight trajectories. Both of
these assumptions, however, are inconsistent with either the
predictions of the EWD-model or the currently reconstructed
picture of the mass-transfer in AE~Aqr.

   \section{Summary and discussion}\label{summary}

Application of the EWD-model to AE~Aqr shows that the high-energy
emission of the system is dominated by the radiative losses of TeV
electrons accelerated in the magnetosphere of the primary. The
energy of these electrons is converted into the TeV photons mainly
due to the inverse Compton scattering. The luminosity of the TeV
source depends on the optical state of the system and lies within
the interval $(3-500) \times 10^{27}\,{\rm erg\,s^{-1}}$. Therefore,
the expected maximum flux of TeV radiation from AE~Aqr (during the
strongest optical flares) within our model is limited to $\la 2
\times 10^{-13}\,{\rm cm^{-2}\,s^{-1}}$. This is below the threshold
of detectors used in all TeV observations of the system reported so
far and therefore, a lack of success in searching for the
high-energy emission of AE~Aqr by the Whipple group proves to be
quite understandable.

On the other hand, the origin of TeV emission detected by the
Potchefstroom and Durham groups within our approach appears to be
a miracle. If this radiation is indeed emitted by AE~Aqr one has
to assume that an acceleration mechanism, which is more powerful
than that investigated in this paper, operates in the system. What
kind of a mechanism could it be\,?

A lack of a disk in the system forces us to reject a number of
previously suggested acceleration scenarios. In particular, the
unipolar inductor model of Cheng \& Ruderman
(\cite{Cheng-Ruderman-1991}) and the dead disk model of Michel \&
Dessler (\cite{Michel-Dessler-1981}) are not applicable in the
light of current view on the mass-transfer process in AE~Aqr.

The statistical acceleration mechanisms operating inside the
magnetosphere of the primary are not effective due to a relatively
small scale of the system. As shown by Kuijpers et~al.
(\cite{Kuijpers-etal-1997}), the typical Lorentz factor of
electrons accelerating by the magnetic pumping in the
magnetosphere of the white dwarf is about 100. This indicates that
the statistical mechanisms of acceleration can be helpful for the
interpretation of the radio emission of the system but their
contribution to the very high energy $\gamma$-rays is negligible.

One can also envisage a situation in which the system acts as an
unipolar inductor due to the interaction between the magnetic
field of the primary and the normal companion, which is partly
situated inside the light cylinder of the white dwarf. This
interaction leads to a perturbation of the magnetospheric field
broadening the area of the hot spots at the magnetic pole regions
by a factor of $(R_{\rm lc}/R_{\rm L1})^{1/2} \simeq 1.25$ and
creating an electric potential in the region of interaction (here
$R_{\rm L1}$ is the distance from the white dwarf to the L1
point). The rate of energy release due to this interaction is
limited to
 \be\label{l1-2}
 L_{1-2} \la  \frac{Q_2 c B^2(R_{\rm L1})}{8 \pi} \simeq 10^{33}\
 \mu_{34}^2\ R_{11}^{-6}\ \left(\frac{R_2}{7\times
 10^{10}}\right)^2\ {\rm erg\,s^{-1}},
 \ee
where $Q_2$ is the effective cross section of the interaction
between the normal companion of the radius $R_2$ and the magnetic
field of the primary. As seen from Eq.~(\ref{l1-2}), the total
power of this interaction is at least a factor of 6 smaller than
the spin-down power of the primary (assuming $\mu \sim
10^{32}\,{\rm G\,cm^3}$ one finds $L_{1-2} \sim 10^{-5}\,L_{\rm
sd}$). This indicates that the contribution of this interaction to
the flux of relativistic particles cannot exceed that of the
pulsar-like acceleration mechanism presented in this paper.
Furthermore, as shown by Rafikov et~al.
(\cite{Rafikov-etal-1999}), the energy released in the process of
the interaction between the magnetic field and a diamagnetic body
moving through the field is converted mainly to the thermal energy
of the body and the energy of Alfv\'en waves generated in this
process.

An effective acceleration of charged particles can occur outside the
magnetosphere in a region of interaction between the relativistic
wind ejected by the white dwarf and the material surrounding the
system. This interaction leads to a formation of a shock at a
distance (see e.g. de~Jager \& Harding \cite{de-Jager-Harding-1992},
and references therein)
 \be
r_{\rm s} \simeq 10^{17}\ \rho_{-24}^{-1/2} V_6^{-1} L_{34}^{1/2}
{\rm cm},
 \ee
where $\rho_{-24}$ and $V_6$ are the density and the thermal
velocity of the material surrounding the system expressed in units
of $10^{-24}\,{\rm g\,cm^{-3}}$ and $10^6\,{\rm cm\,s^{-1}}$, and
$L_{34}=L_{\rm sd}/10^{34}\,{\rm erg\,s^{-1}}$. The energy of
particles accelerated in the shock can be as high as
 \be
E_{\rm sh, max} \sim 10^{13}\ B_{-6}\ (l_0/r_{\rm s})\ {\rm eV},
 \ee
where $B_{-6}$ is the magnetic field strength in the shock region
expressed in units of $10^{-6}$ (this normalization is derived by
equating the magnetic field energy density, $B^2/8\pi$, to the
energy density of the background material, $(1/2) \rho v_{\rm
s}^2$). The radiative losses of these particles under these
conditions are dominated by the synchrotron emission and the energy
of the corresponding photons lies within the region of $\sim
200$\,MeV. Furthermore, the radiation time of these electrons is
close to $10^6$\,yr. Therefore, the energy radiated by the electrons
on a scale of $r_{\rm s}$ is small. This indicates that a
contribution of the shock into TeV emission of the system is almost
negligible.

A situation in which the white dwarf could appear as an ejector of
very high energy particles with a rate $\sim 10^{34}\,{\rm
erg\,s^{-1}}$ has been discussed by Meintjes \& de~Jager
(\cite{Meintjes-de-Jager-2000}). A key assumption of their approach
is that the efficiency of the drag-driven propeller action by the
white dwarf is too low for the most dense blobs to be expelled from
the system. These blobs, therefore, are able to reach the
circularization radius (which for the parameters of AE~Aqr is
$r_{\rm circ} \sim 2 \times 10^{10}$\,cm) and to form a clumpy disk
surrounding the magnetosphere of the white dwarf. A mixing of the
disk material with the magnetic field, which is assumed to be
governed by a turbulent diffusion and the Kelvin-Helmholtz
instability, leads to a perturbation and reconnection of the field
lines. The energy releases in the corresponding current sheets is
assumed to be transferred into the energy of particles accelerated
in these regions (for a discussion see also Meintjes \& Venter
\cite{Meintjes-Venter-2005}).

The energy of accelerated particles and the rate of their ejection
evaluated within the MHD-propeller model are high enough for the
flux of TeV emission to be comparable to that reported by
Potchefstroom and Durham groups. Furthermore, this model is also
helpful for an interpretation of the system radio and mid-infrared
emission (Meintjes \& Venter \cite{Meintjes-Venter-2005}). The
reasons for a lack of success in searching for the high-energy
emission of AE~Aqr by the Whipple group in this case might be a
relatively low probability to detect infrequent TeV activity of
the star (Lang et~al. \cite{Lang-etal-1998}), or/and the energy
dependence of the image selection criteria, used by the Whipple
group in their data analysis procedure, which may lead to a
significant loss of $\gamma$-ray events in case of sources with
spectra harder that that of Crab Nebular (for a discussion see
Bhat et~al. \cite{Bhat-etal-1998}).

At the same time, a conclusion that MHD-propeller model provides
us with a comprehensive explanation of the energy release process
in AE~Aqr seems to be rather premature. The model is built around
an assumption about a very high rate of plasma penetration into
the magnetic field of the white dwarf. In particular, the
diffusion coefficient is normalized by Meintjes \& Venter
(\cite{Meintjes-Venter-2005}) to its absolute maximum value
$10^{16}\,{\rm cm^2\,s^{-1}}$, which for the conditions of
interest is 10 orders of magnitude larger than the value of the
Bohm diffusion coefficient. Such a situation is unique in
astrophysical objects. In particular, it has never been observed
in solar flares, interplanetary space (see e.g. Priest \& Forbes
\cite{Priest-Forbes-2000}), and accretion-driven compact stars
(Frank et~al. \cite{Frank-etal-1985}). Furthermore, the time of
blob diffusion into the magnetic field is evaluated to $\sim
30$\,s. This is a factor of 2 smaller than the free-fall time,
$t_{\rm ff} = r^{3/2}/(2 GM_{\rm wd})^{1/2}$ at a distance of
closest approach of the blobs to the white dwarf ($\sim
10^{10}$\,cm). This indicates that the rate of diffusion evaluated
within the MHD-propeller model significantly exceeds the rate of
plasma penetration into the magnetic field of a neutron star
evaluated by Arons \& Lea (\cite{Arons-Lea-1976}) for the case of
spherical accretion and by Ghosh \& Lamb (\cite{Ghosh-Lamb-1979})
for the case of disk accretion. Finally, the diffusion time of the
blobs into the magnetic field within this approach is
significantly smaller than the characteristic time of drag
interaction between the blobs and the magnetic field, which
according to Wynn et~al. (\cite{Wynn-etal-1997}) is $t_{\rm drag}
\gg t_{\rm ff}$. This rises a question about the structure of the
H$\alpha$ Doppler tomogram expected within the MHD-propeller
model. A simulation of this tomogram within the MHD-propeller
model may help us to answer the question why the contribution of
the clumpy disk into the observed tomogram is negligibly small.

The above mentioned items suggest that the theoretical development
of the MHD-propeller model remains sofar work in progress.
Nevertheless, the basic prediction of this model, namely, a flux of
TeV emission from AE~Aqr over the level of $10^{-12}\,{\rm
cm^{-2}\,s^{-1}}$ can be observationally tested with the MAGIC
(Lorenz \cite{Lorenz-2004}), VERITAS (Weekes et~al.
\cite{Weekes-etal-2002}) and  HESS (Hinton \cite{Hinton-2004})
experiments. The corresponding observations will allow us to choose
between the MHD-propeller model and EWD-model which suggests that
the TeV flux of the system is below the threshold of detectors used
in the above mentioned high-energy experiments.

\begin{acknowledgements}
   We thank Anatoli Tsygan for fruitful discussion and an anonymous
   referee for careful reading of the manuscript and useful comments.
   Nazar Ikhsanov acknowledges the support of the Alexander von Humboldt
   Foundation within the Follow-up Program and the European Commission
   under the Marie Curie Incoming Fellowship Program. The work was
   partly supported by the Korea Science and Engineering Foundation
   under the grant R01-2004-000-1005-0, Russian Foundation of Basic
   Research under the grant 03-02-17223a and the Russian State
   Scientific and Technical Program ``Astronomy''.
\end{acknowledgements}

\end{document}